\documentclass[useAMS,usenatbib]{mnras}
\usepackage{graphicx}
\usepackage{amssymb}

\newcommand{\kms}{km\,s$^{-1}$}
\newcommand{\ha}{\textrm{H}$\alpha$}
\newcommand{\FeH}{\textrm{[Fe/H]}}

\title[A dwarf galaxy discovery near the LV dwarf]{Serendipitous discovery of 
a faint dwarf galaxy near a Local Volume dwarf
\thanks{Based on observations made with the NASA/ESA Hubble
Space Telescope, program GO-13442, with data archive at the Space
Telescope Science Institute. STScI is operated by the Association of 
Universities for Research in Astronomy, Inc. under NASA
contract NAS 5--26555.}
}
\author[L. N. Makarova et al.]{
L. N. Makarova$^{1}$\thanks{E-mail: lidia@sao.ru}, 
D. I. Makarov$^{1}$,
A. V. Antipova$^{1}$,
I. D. Karachentsev$^{1}$,
R. B. Tully$^{2}$\\
$^{1}$Special Astrophysical Observatory, Nizhniy Arkhyz, 
Karachai-Cherkessia 369167, Russia\\
$^{2}$Institute for Astronomy, University of Hawaii, 2680 Woodlawn Drive, 
HI 96822, USA
}
\begin{document}

\date{Accepted XXX. Received XXX; in original form XXX}

\pagerange{\pageref{firstpage}--\pageref{lastpage}} \pubyear{XXX}

\maketitle

\label{firstpage}

\begin{abstract}
A faint dwarf irregular galaxy has been discovered in the HST/ACS 
field of LV\,J1157+5638. The galaxy is resolved into individual stars, 
including the brightest magnitude of the red giant branch. 
The dwarf is very likely a physical satellite of LV\,J1157+5638.
The distance modulus of LV\,J1157+5638 using the tip of 
the red giant branch (TRGB) distance indicator is $29.82\pm0.09$~mag ($D = 9.22\pm0.38$~Mpc).
The TRGB distance modulus of LV\,J1157+5638\,sat
is $29.76\pm0.11$~mag ($D = 8.95\pm0.42$~Mpc).
The distances to the two galaxies are consistent within the uncertainties.
The projected separation between them is only 3.9 kpc.
LV\,J1157+5638 has a total absolute V-magnitude of $-13.26\pm0.10$ and 
linear Holmberg diameter of 1.36 kpc, whereas its faint satellite 
LV\,J1157+5638\,sat has  $M_V = -9.38\pm0.13$ mag and 
Holmberg diameter of 0.37 kpc. Such a faint dwarf was discovered for 
the first time beyond the nearest 4 Mpc from us.
The presence of main sequence stars in both galaxies unambiguously
indicates the classification of the objects as dwarf irregulars (dIrrs) with
recent or ongoing star formation events in both galaxies.
\end{abstract}


\section{Introduction}

\begin{figure*}
\centering
\begin{tabular}{p{0.5\textwidth}@{}p{0.4\textwidth}}
 \includegraphics[width=0.5\textwidth]{fig1a.ps} &
 \raisebox{64mm}{
   \begin{tabular}{p{0.4\textwidth}}
   \includegraphics[width=0.4\textwidth]{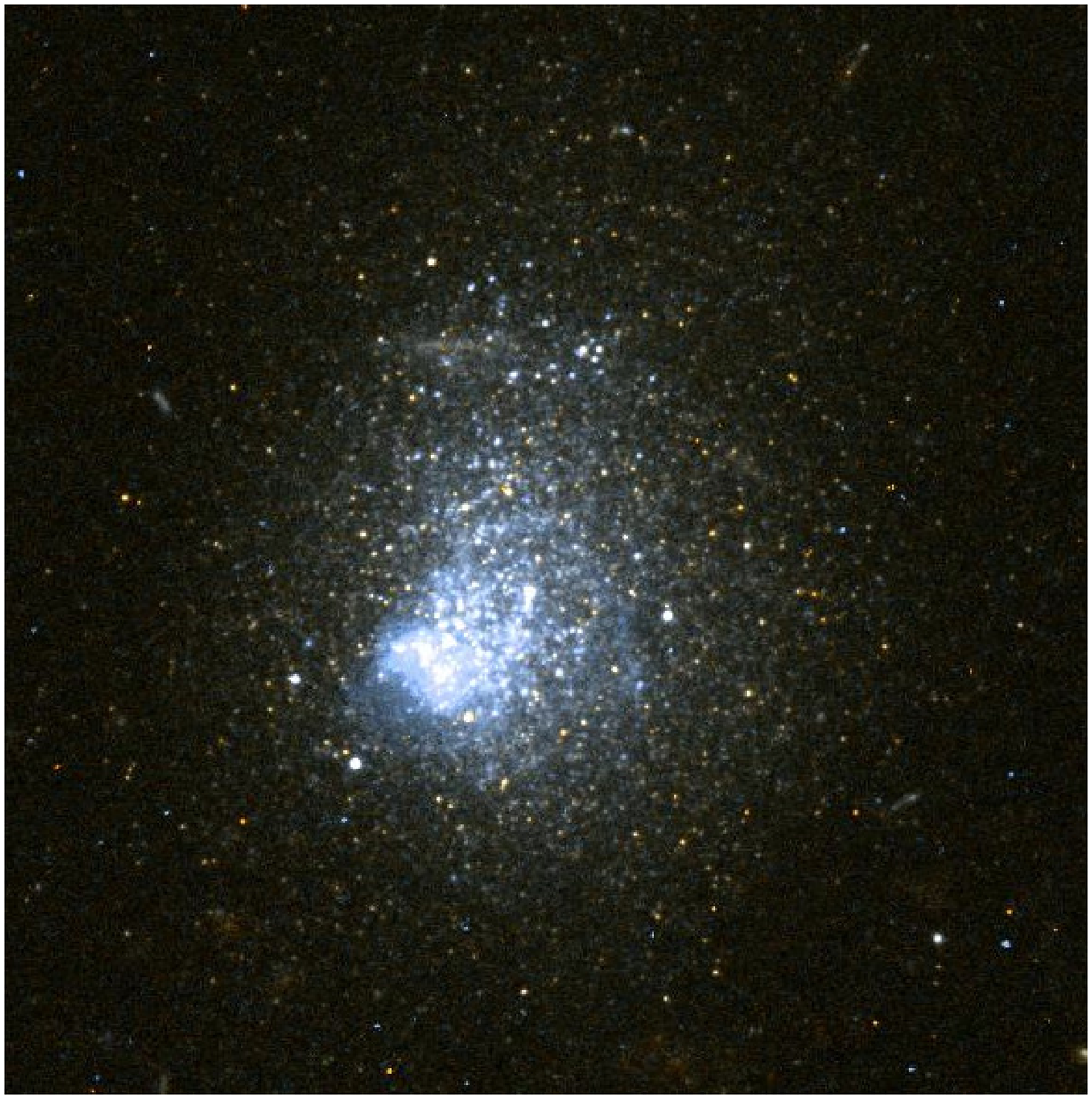} \\
   \includegraphics[width=0.4\textwidth]{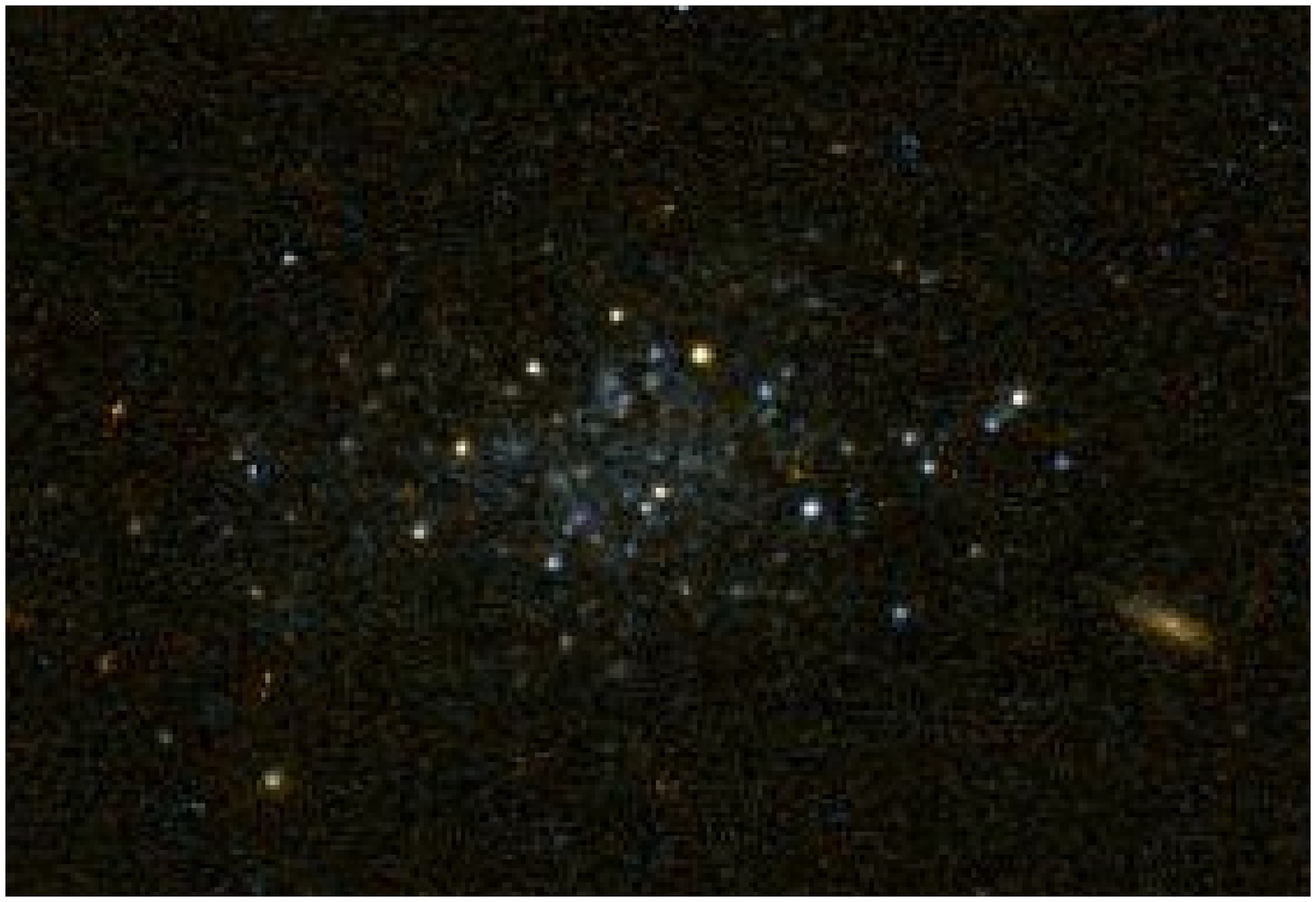}
   \end{tabular}
 }
\end{tabular}
\caption{\textit{HST}/ACS combined distortion-corrected  mosaic image 
of the LV\,J1157+5638 field in the \textit{F606W} filter (left panel). 
The image size is $1.2\times1.8$ arcmin. 
Enlarged combined $F606W+F814W$ images of LV\,J1157+5638
and the new dwarf satellite are shown at the right panel.
}
\label{fig:ima}
\end{figure*}

In recent years, the search for previously unknown dwarf galaxies has 
attracted considerable attention.
This interest is associated with a number of unsolved, but well-known 
problems with modern cosmological $\Lambda$CDM theory. The problem of 
`lost satellites' is one of them. The luminosity 
function of the galaxies in the Local Volume contains about 
an order of magnitude fewer dwarfs than predicted by 
the $\Lambda$CDM theory \citep{klypin99,klypinetal2015}.
Consequently, the search for missing satellites is especially important,
and this search has been quite successful in recent years.
Particularly rich families of faint satellites were discovered around the two giant 
galaxies of the Local Group: Andromeda and the Milky Way \citep{laevens2015,homma2016,bechtol2015,koposov2015}.
The discovery of very faint nearby dwarf galaxies 
makes it possible to clarify the structure and dynamics of our Local Group, 
which in turn is necessary for constructing a model of the  formation and evolution
of the Local Group.

For more distant galaxy groups we naturally lose more faint satellites.
Nevertheless, recent observations on ground-based telescopes, and
follow-up observations with the Hubble Space Telescope, substantially reduce
these gaps. Such a survey was performed for the M\,81 group \citep{chib2009}, 
where 12 new dwarf galaxies were found. Recently, searches for faint 
galaxies in the Centaurus\,A group were successfully carried out \citep{crno2016,muller2017}, 
and these surveys led to the discovery of more than 60 new dwarf galaxy candidates.
Even in more distant groups like, for example, the M\,101 group at the
distance of 7.2 Mpc \citep{lee2012}, observational surveys bring the discovery of 
new dwarf satellites \citep{java2016,mullerm101,danieli2017}. \citet{park2017} also
recently reported the discovery of 22 dwarf members of the group around NGC 2784 
(at the distance about 9.8 Mpc).

However, not only groups around giant galaxies should be taken into account.
\citet{km2008} in the framework of a binary galaxy study in the Local Supercluster 
drew attention to the existence of a large number of systems consisting exclusively 
of dwarf galaxies. 
A similar claim was made by \citet{tully2006}.
\citet{mu2012} have compiled a sample of similar groups formed 
exclusively of dwarf galaxies with $ M_K> -19 $ in the $ K $ filter. The majority of
the list consists of galaxy pairs, and the most populated system contains six members. 
Dwarf galaxy groups account for about 5\% of all groups in the Local Supercluster \citep{mu2012}.
Taking into account the selection effects, the total number of multiple dwarf systems 
should be at least 5-6 times greater. The authors show that groups of dwarf galaxies 
are located in low density regions and evolve without the influence of massive neighbours.

Despite difficulties in finding of faint satellites of dwarf galaxies, a number of 
discoveries have recently been made \citep{chen2013,crno2014,carlin2016,smer2017}.
Although targeted hunts give us a substantial increase in the number of dwarf 
galaxies of the Local Volume, new objects can also be found in 
the analysis of serendipitously observed areas of sky.
In this paper, we report the discovery of an extremely small galaxy 
located near the LV\,J1157+5638 dwarf galaxy. The object was detected 
by us during the data analysis of images from
HST/ACS program SNAP--13442.

\section{Observations and data reduction}

LV\,J1157+5638 was observed on October 18, 2013 with HST/ACS 
in the course of the SNAP project 13442 (PI: R.B.Tully), that was undertaken to acquire
accurate photometric distances of a sample of Local Volume 
galaxies. With accurate distances, we generate three-dimensional
maps of the distribution of galaxies and decouple the expansion 
and peculiar components of line-of-sight velocities.
Dithered images were obtained in the $F606W$ and $F814W$
filters with the exposures summing to 1100 s in each band.
Initial reductions used the default HST pipeline.
The $F606W$ image of the LV\,J1157+5638 field is shown in Fig.~\ref{fig:ima}.
This compact galaxy is very well resolved into individual stars. 
It is easily distinguished in the upper part of the ACS image. 
This region can be seen in detail on the enlarged image at the upper right panel.
The off-centre knot of blue brighter stars is well resolved into individual stars, indicating
ongoing star formation in LV\,J1157+5638. A newly discovered galaxy is situated about 1.5 arcmin 
to the south of LV\,J1157+5638 in the lower right corner of the image. It is shown at the lower right panel, 
and also visibly resolved into individual stars.

We use the ACS module of the {\sc DOLPHOT} software 
package\footnote{\url{http://americano.dolphinsim.com/dolphot/}}
by A. Dolphin for photometry of resolved stars.
The HST data quality files were used to mask bad pixels. 
{\sc DOLPHOT} parameters were set and the photometry was performed as recommended 
in the {\sc DOLPHOT} User's Guide for ACS/WFC data.
Only stars of good photometric quality were used in the
analysis, following the recommendations given in the manual. 
We have selected the stars with signal-to-noise (S/N) of at least 
five in both filters and $\vert sharp \vert \le 0.3$.
The resulting colour-magnitude diagrams (CMD) of the LV\,J1157+5638 and its satellite are presented in 
Fig.~\ref{fig:cmd}. We call this satellite LV\,J1157+5638\,sat in this paper.

We use artificial star tests incorporated within {\sc DOLPHOT} 
to estimate the photometric errors, blending and 
incompleteness in the crowded fields of nearby resolved galaxies.
A significant number of artificial stars were generated 
in the appropriate magnitude and colour range so that
the distribution of the recovered magnitudes is adequately sampled. 
According to the artificial star experiments, the 50\% completeness level is
appearing at $F814W$ $\simeq$27.2~mag and at $F606W$ $\simeq$28.4~mag. 

LV\,J1157+5638\,sat was observed on June 10, 2017 at the 6-meter BTA telescope (SAO RAS)
with the multi-mode focal reducer of the BTA SCORPIO in the long slit mode 
with a grism VPHG1800R \citep{af2005}, providing reciprocal dispersion 0.52 $\AA$/pix
and spectral range 6100--7100 $\AA$. The spectral resolution is 2.5 $\AA$.
The total exposure time was 1800 seconds. Unfortunately, 
\ha{}-emission was not seen, thus, we could not measure the redshift.

\section{The colour-magnitude diagrams}

\begin{figure*}
\includegraphics[height=10cm]{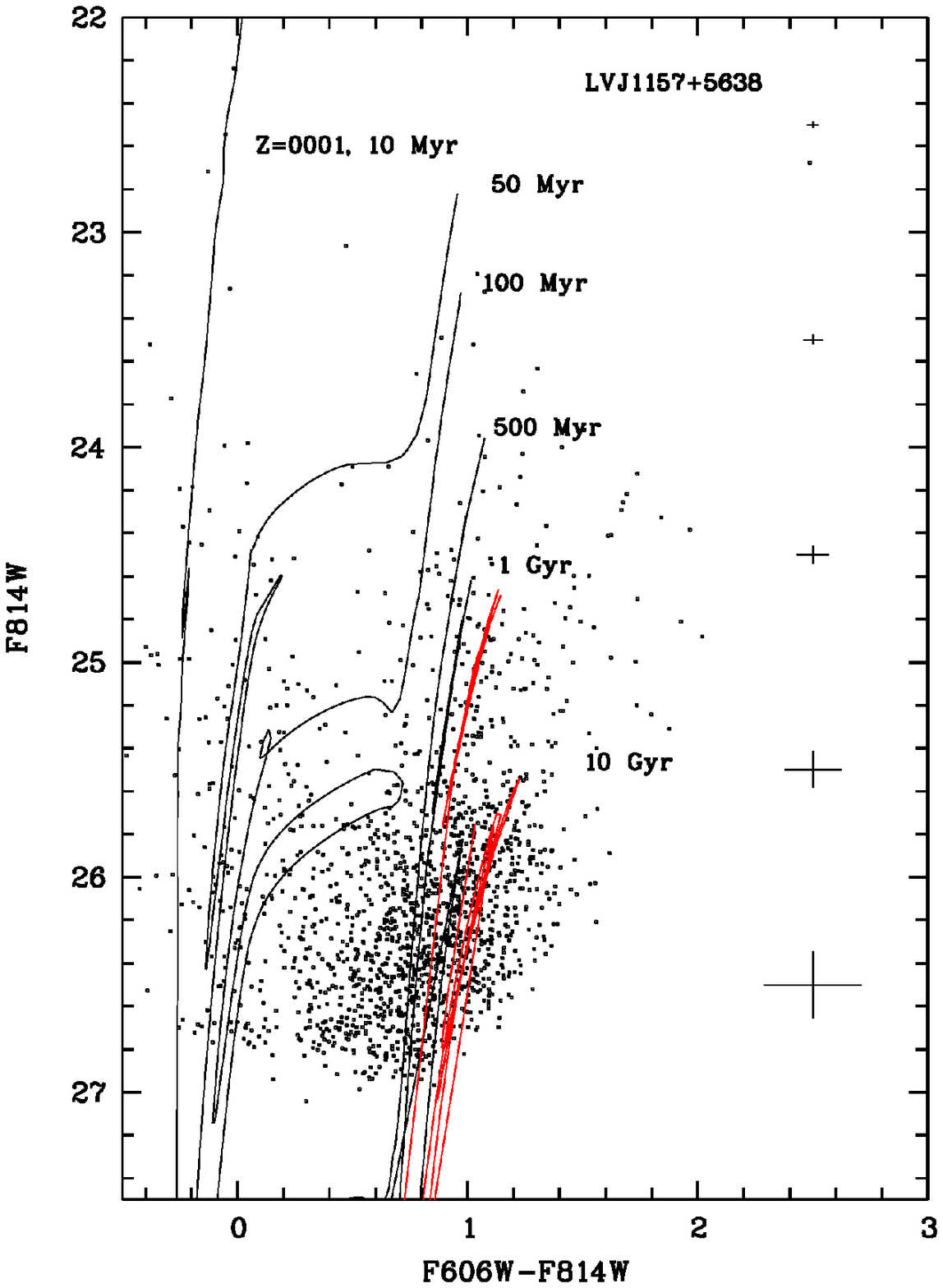}
\includegraphics[height=10cm]{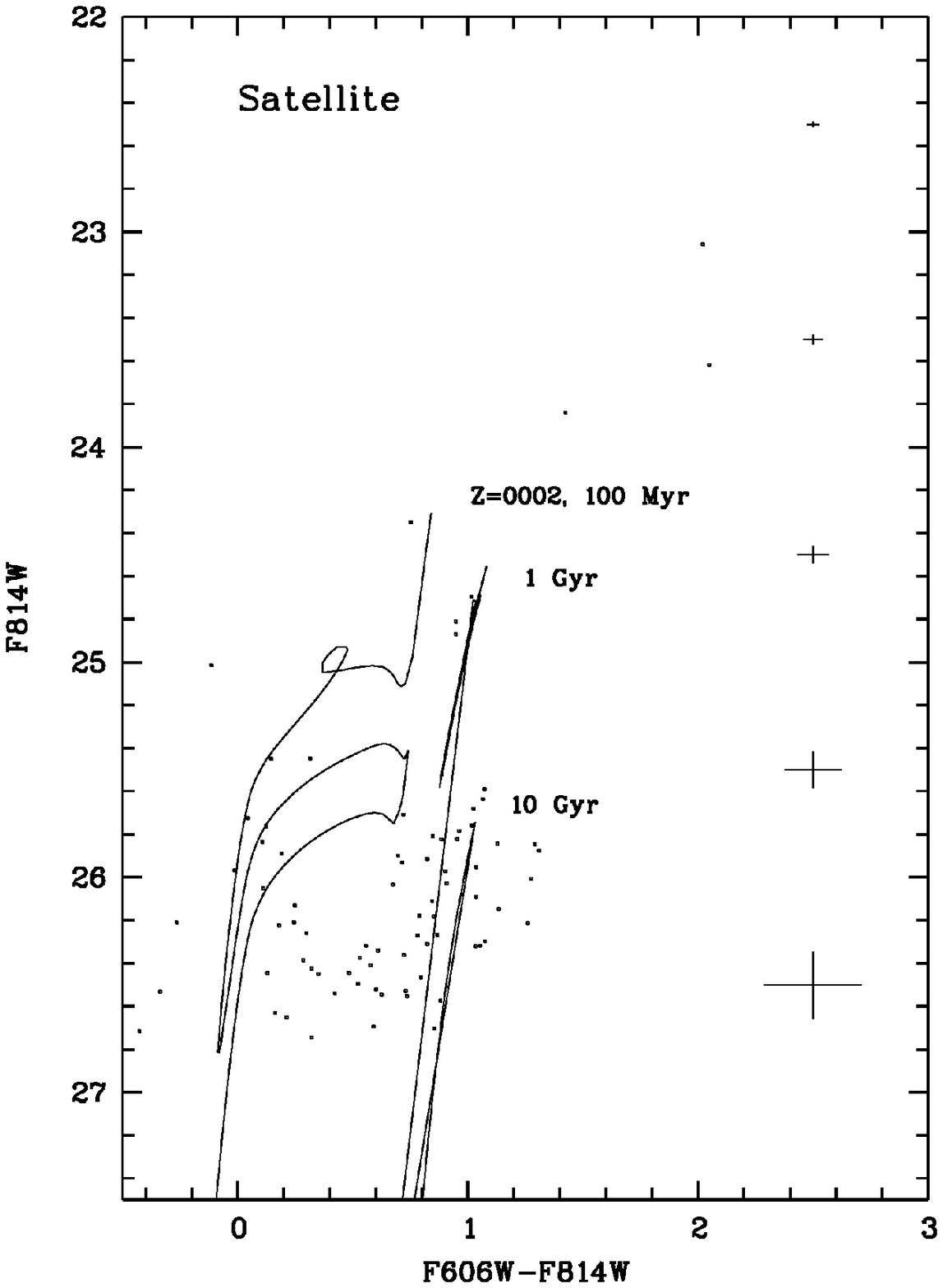}
\caption{LV\,J1157+5638 and LV\,J1157+5638\,sat colour-magnitude diagrams. 
Photometric errors are indicated by the bars at the right in the CMD.
Padova theoretical stellar isochrones \citep{marigo2017} of 
the different ages and metallicities are overplotted. 
The metallicity of the shown isochrones are: left panel, black Z=0001
($\FeH = -2.36$); left panel, red Z=0.0004 ($\FeH = -1.74$); right panel Z=0.0002
($\FeH = -2.05$).
}
\label{fig:cmd}
\end{figure*}

Two colour-magnitude diagrams are presented in the Fig.~\ref{fig:cmd}. In the left panel 
we show stellar populations measured at the ACS/WFC field within the body of LV\,J1157+5638, and
at the right panel are the stars within the tiny satellite LV\,J1157+5638\,sat.
Even the CMD of the `main' dwarf looks sparsely populated. We can see
upper main sequence at $(F606W-F814W)\leq0.4$, red supergiant plus upper AGB 
(asymptotic giant branch) stars at $F814W\leq25.7$ mag, and the rest are more
abundant RGB (red giant branch). Only about 80 stars were resolved in the tiny 
satellite LV\,J1157+5638\,sat. Nevertheless, clear signs of the upper main sequence
at $(F606W-F814W)\leq0.4$ are presented, and the RGB is well represented.  Among
brighter stars ($F814W\leq25$ mag) we can distinguish a few red supergiants,
but the main sequence only rises to roughly match the level of the TRGB at $F814W$. 
It is obvious that we still have
populations in this dwarf which are not fully resolved in the body of the galaxy
with these exposures, and deeper images are needed. At the same time,
the presence of main sequence stars in both galaxies unambiguously
indicate, that we can classify the objects as dwarf irregulars (dIrrs).
Therefore, we can expect recent or ongoing star formation events in the galaxies.
According to the GALEX and \ha{} data (see Table~\ref{table:general}, where the general
parameters and results are indicated), LV\,J1157+5638 has
sufficient ongoing star formation, whereas its satellite is too faint to evince
ongoing star formation activity with the data available.

\section{Distance measurement and star formation}
We have determined a photometric TRGB distance of both LV\,J1157+5638 dwarf 
galaxy and its expected satellite with
our {\sc trgbtool} program which uses a maximum-likelihood algorithm 
to obtain the magnitude of TRGB from the stellar
luminosity function \citep{makarov06}. 
The measured TRGB magnitude of LV\,J1157+5638 is $F814W_{\rm TRGB} = 25.74\pm0.07$ mag 
in the ACS instrumental system. 
A photometric distance of LV\,J1157+5638 dwarf galaxy was also estimated
with the same data in the Extragalactic Distance Database (EDD)\footnote{\url{http://edd.ifa.hawaii.edu/}}.
The TRGB magnitude given there is $F814W_{\rm TRGB} = 25.66\pm0.13$ mag.
The both estimations are consistent within 1$\sigma$ uncertainty.
Using the calibration for the TRGB distance indicator by 
\citet{rizzietal07} and the Galactic extinction $E(B-V) = 0.017$ from \citet{schlafly}, 
we derived the true distance modulus for LV\,J1157+5638:
$29.82\pm0.09$ mag ($D = 9.22\pm0.38$ Mpc).
The CMD of the LV\,J1157+5638\,sat is poorly populated, and the galaxy is distant
which makes uncertainties of the distance
estimation large. 
Nevertheless, the {\sc trgbtool} program is working quite well even in this case.
The measured TRGB magnitude of LV\,J1157+5638\,sat is $F814W_{\rm TRGB} = 25.68\pm0.09$ mag.
We obtained the distance modulus for this dwarf $29.76\pm0.11$ mag ($D = 8.95\pm0.42$ Mpc).
The distances to the both galaxies are consistent within the uncertainties.
Therefore, we can claim, that the considered dwarfs are very likely a physical pair with 
the projected separation 3.9 kpc between them. 

Since the red giant branch is clearly visible in both our galaxies, 
we can estimate the average metallicity of these stars. 
According to \citet{lee}:
$$\FeH = -12.64 + 12.6(V-I)_{-3.5} - 3.3(V-I)^2_{-3.5}, $$
where $(V-I)_{-3.5}$ is the colour of RGB stars at the level of 0.5 mag fainter than TRGB value. 
We calculated $\FeH = -2.30\pm0.07$ dex and $\FeH = -2.08\pm0.10$ dex for the main galaxy and its satellite, respectively.

The isochrones of these metallicities were superimposed on the CMDs of the studied dwarfs 
(see Fig.~\ref{fig:cmd}), so that we can approximately estimate the age of the resolved stars.
In the LV\,J1157+5638 galaxy, we can assume the presence of a small number of stars $\sim$10 Myr 
old, which indicates an evident ongoing star formation, as can be seen also from \ha{} data 
(see section 3). At the same time, the upper main sequence of the dwarf is not densely populated, 
so we cannot expect a recent intense burst of star formation.
Relatively young stars of the age of 50--100 Myr are also present in the galaxy. In addition, 
we can assume from the theoretical isochrones, that the age of the resolved red giants can be 
from 1 to 10--13 Gyr, i.e. the galaxy most likely includes the oldest RGB stars.
As can be seen from the figure, we can not exclude somewhat higher metallicity of the RGB stars,
$\FeH = -1.74$. However, photometric errors play a significant role in this 
part of the CMD, that it is difficult to make certain conclusions.

Apparently, there are no stars younger 100 Myr in the tiny satellite (see the right panel of 
Fig.~\ref{fig:cmd}). This agrees with the \ha{} and GALEX data, and indicates 
the absence of ongoing star formation. The isochrones of the metallicity estimated from 
the RGB colour well fit the CMD as a whole. LV\,J1157+5638\,sat also most likely includes 
the oldest RGB stars up to about 13 Gyr old.

\section{Integral and surface photometry}

\begin{figure}
\includegraphics[width=8cm]{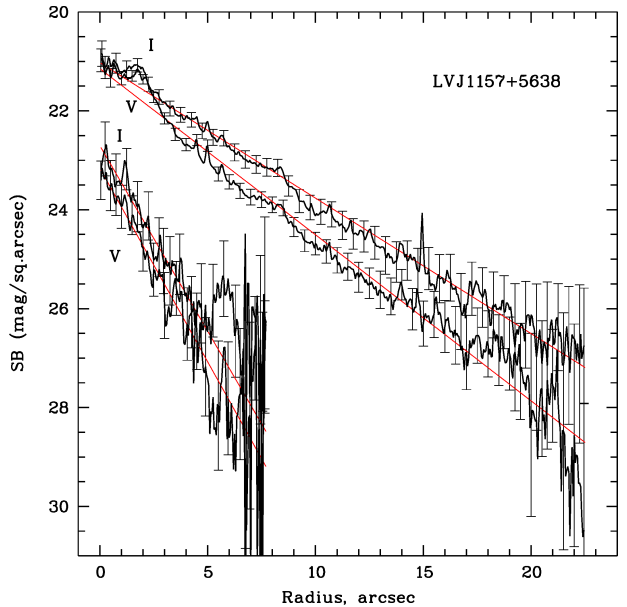}
\caption{
The surface brightness profiles of LV\,J1157+5638 and LV\,J1157+5638\,sat in $V$ and $I$ bands.
Photometric uncertainties are indicated by the vertical bars. Exponential intensity law
fitting is shown with the red solid lines.
}
\label{fig:profile}
\end{figure}

Both LV\,J1157+5638 and LV\,J1157+5638\,sat dwarfs are situated well within the ACS/WFC
field, allowing us to perform integral and surface photometry of the galaxies.
Total and surface photometry was made on the fully processed 
distortion-corrected HST/ACS $F606W$ and $F814W$ images.
Foreground stars were removed from the frames by fitting a first order 
surface in a rectangular pixel-area in the nearest neighbourhood of a star. 
The sky background in the ACS images is insignificant but, to remove
possible slight large scale variations, 
the sky was approximated by a tilted plane, created from 
a two-dimension polynomial, using the least-squares method. The accuracy 
of the sky background determination is about 1 -- 2 \% of the original sky level.
To measure total magnitudes in $F606W$ and $F814W$ bands, 
integrated photometry was performed in increasing circular apertures
from the visual geometric centre to faint outskirts of the galaxies.
The total magnitude was then estimated as the asymptotic value of
the obtained radial growth curve. The measured total magnitudes are 
$V = 16.61\pm0.04$ mag and $I = 16.05\pm0.04$ mag for the main galaxy, and 
$V = 20.43\pm0.06$ mag and $I = 19.75\pm0.06$ mag for the satellite. The estimated errors 
include the photometry and sky background uncertainties, as well as
the transformation errors from instrumental ACS magnitudes to the standard
$V$ and $I$ magnitudes \citep{sirianni}. The respective absolute magnitudes
of  LV\,J1157+5638 are $M_V = -13.26\pm0.10$ and $M_I = -13.80\pm0.10$, taking into account
Galactic extinction \citep{schlafly}, and the distance modulus from the
present paper (see above). The absolute magnitudes for 
LV\,J1157+5638\,sat are $M_V = -9.38\pm0.13$ and $M_I = -10.04\pm0.13$.
Azimuthally averaged surface brightness profiles of LV\,J1157+5638 
and LV\,J1157+5638\,sat were obtained by differentiating the galaxy growth curves with
respect to the radius. The resulting profiles in $V$ and $I$ bands
are displayed in Fig.~\ref{fig:profile}. They are calculated to the level of 
about 27.5 mag\,arcsec$^{-2}$ in $I$ band and about 29.5~mag\,arcsec$^{-2}$ in $V$ band.
The surface brightness of LV\,J1157+5638\,sat is very low, resulting the quite noisy profiles. 

The calculated profiles of the main galaxy and its satellite were fitted
by an exponential intensity law of brightness distribution, which is generally
appropriate for both dwarf irregular and spheroidal galaxies:
$$ \mu(r) = \mu_0+1.086\times(r/h), $$
where $\mu_0$ is the central surface brightness and $h$ is the
exponential scale length.
The galaxies follow this distribution law quite well, excluding the hump at 1-3 arcsec 
of the radius, caused by a noticeable star-forming region in the main galaxy.
Unweighted exponential fits to the surface brightness profiles were obtained 
by linear regression. The derived central surface brightness and 
the exponential scale lengths are given in Table~\ref{table:general}.
The uncertainties are formal fitting errors.

\section{Discussion and concluding remarks}
\begin{figure*}
\includegraphics[width=\textwidth]{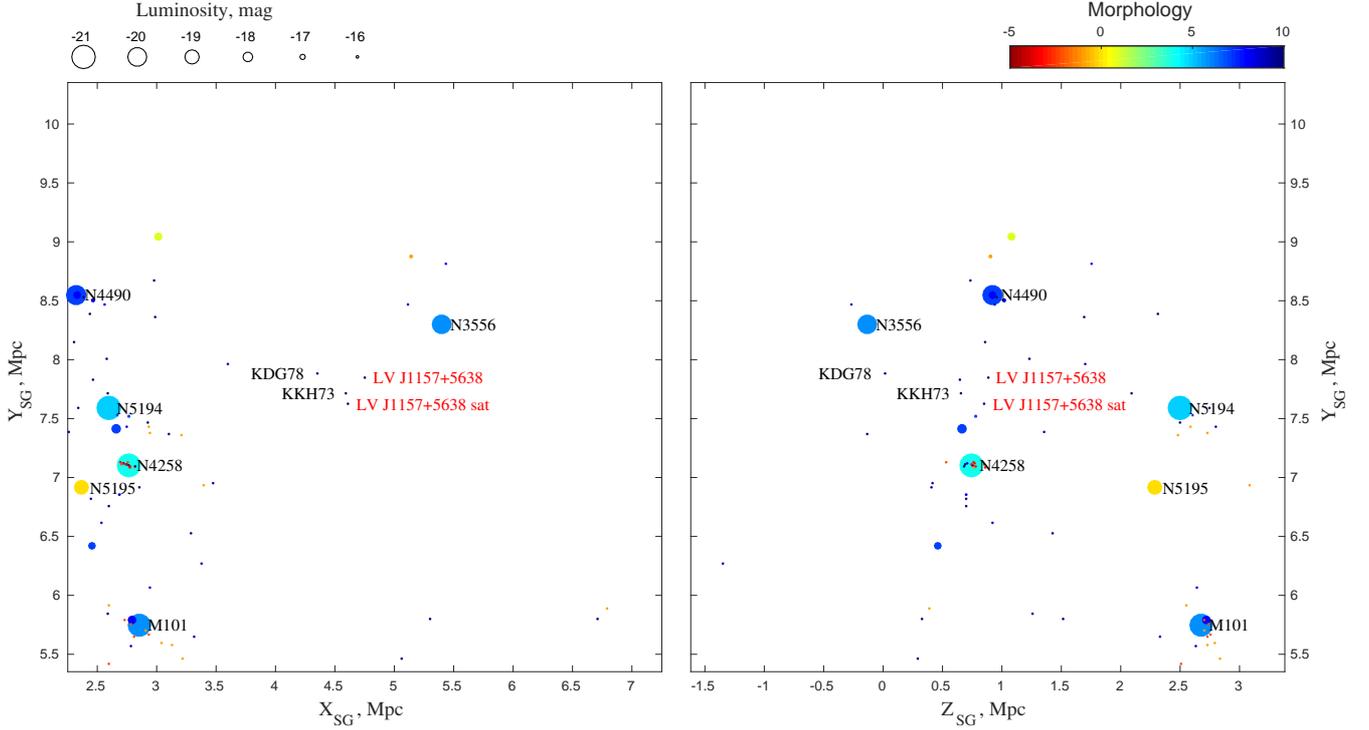}
\caption{A panorama of the LV\,J1157+5638 neighbourhood in the supergalactic coordinates.
The figure shows the projection of galaxies in a cube of $\pm 2.5$~Mpc size .
The left panel is a projection on the supergalactic plane XY, while the right 
panel is the ZY view of the distribution of galaxies. The colour of a dot represents 
the morphology of the galaxy according to the colour bar.
The size of a galaxy corresponds to its luminosity as shown in the legend panel.
}
\label{fig:structure}
\end{figure*}

\begin{figure*}
\includegraphics[width=10cm]{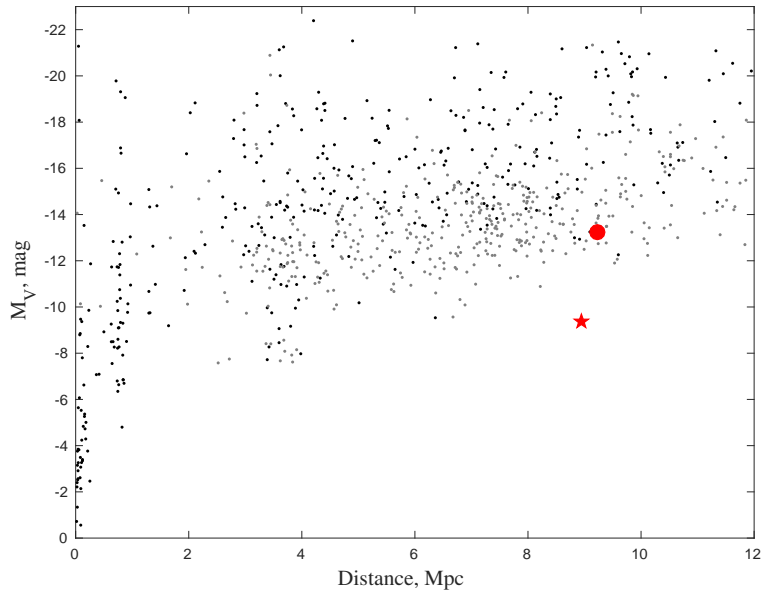}
\caption{A relation between linear distance and total absolute V-magnitude for
the Local Volume galaxies. The data are taken from the HyperLeda database.
Black dots are represent original measurements, and grey dots are the magnitudes originally
measured in B and translated to the V-magnitudes according to the mean colours 
from \citet{sharina2013}. LV\,J1157+5638 is shown with red circle and LV\,J1157+5638\,sat 
with red star.
}
\label{fig:dist-mag}
\end{figure*}
We discover a new faint dwarf irregular galaxy, detected in the HST/ACS 
images. The galaxy is resolved into individual stars, including the RGB, which allowed us 
to measure the TRGB distance to this galaxy.
This dwarf is very likely a physical companion of LV\,J1157+5638.
Thus, we were able to detect
a satellite of a dwarf galaxy. The structure of the neighbourhood of our objects 
is demonstrated in the Fig.~\ref{fig:structure}. It is obvious, that LV\,J1157+5638
is situated far away from any giant galaxies and their satellite families.
The closest neighbour, dwarf irregular galaxy KKH~73, is situated at the projected distance of 83 arcmin (220~kpc) from LV\,J1157+5638. 
The second nearest galaxy, dwarf irregular  KDG~78, is located at the projected distance of 352 arcmin (940~kpc) 
from LV\,J1157+5638 (see Fig.~\ref{fig:structure}).
Unfortunately, both neighbours, KKH~73 and KDG~78, do not have photometric distance estimations. 
Their heliocentric radial velocities $V_h(\textrm{KKH\,73}) = 596 \pm 6$~\kms{} and $V_h(\textrm{KDG\,78}) = 574.8 \pm 1.7$~\kms{}, 
from the LV database\footnote{\url{http://www.sao.ru/lv/lvgdb}} 
exceed the radial velocity of $V_h(\textrm{LV\,J1157+5638})=416.3\pm1.4$ over 150~\kms{}. 
It is highly unlikely that they form a physically bounded system.

According to \citet{klypinetal2015}, the LV galaxy sample is complete up to $M_B\sim-14$ mag. 
We can estimate the total number of fainter galaxies using the Schechter luminosity function 
approximation with the parameters $\phi_*=1.25\times10^{-2}h^3$~Mpc$^3$, $\alpha=-1.3$ and 
$M_*=-20.0+5*\log(h)$ 
in $B$ filter, taking into account that $h=0.73$. The absolute magnitude of LV\,J1157+5638\,sat
is approximately $M_B=-8.9$, assuming the average colour of the LV dIrrs to be 
$\langle B-V\rangle=0.48\pm0.2$ \citep{sharina2013}. The resulting expected number of galaxies 
in the Local Volume ($D<10$~Mpc) with luminosities from $M_B=-8$ to $M_B=-14$ is $1830$.
Assuming a random distribution of these galaxies, we can estimate the probability
of random projection of one of these galaxies into a circle with a radius of 1.5 arcmin, 
which is approximately 0.017 per cent. Similarly, a random location of this galaxy 
inside a sphere with a radius of 0.3~Mpc is 4.9 per cent. Thus, there is a very small 
probability of accidental detection of two dwarf galaxies in such a small spatial area,
i.e. the two studied dwarfs are very likely to be physically connected.

General parameters of the galaxies under study are presented in Table~\ref{table:general}. 
Judging by the total and surface photometry data in the table, 
LV\,J1157+5638\,sat is similar to extreme Local Group dwarfs. 
At the same time LV\,J1157+5638\,sat looks quite faint and rather compact and could be 
similar to d1005+68, the satellite of a dwarf galaxy in the M\,81 group discovered 
by \citet{smer2017}.

Fig.~\ref{fig:dist-mag} represent total absolute V-magnitudes of the Local Volume dwarf
galaxies versus their linear distances. The data were extracted from the HyperLeda database 
\citep{makarov2014}.
Here black dots are measurements from original works, and grey dots are the magnitudes originally
measured in B and translated to the V-magnitudes according to the mean colours
of the LV dwarf galaxies of different types from the work of \citet{sharina2013}. LV\,J1157+5638
is shown with red circle and LV\,J1157+5638\,sat with red star. It is interesting
to note, that the Local Group dwarf galaxy family is relatively well studied, a lot of
really faint objects are discovered. A number of known faint dwarf galaxies is 
rapidly decreases with increasing distance. It is obvious, that LV\,J1157+5638\,sat is
extremely faint for its distance. It is highly possible, that most of faint satellites
are still unknown at the distance of 5--10 Mpc.
 
According to \citet{mu2012}, their groups of dwarf galaxies form a continuous sequence 
in the distribution of luminosities and masses with associations of dwarfs discovered by 
\citet{tully2006} in an analysis of the three-dimensional distribution 
of nearby galaxies. 
The dwarf companion LV\,J1157+5638\,sat discovered by us, 
together with its `main' irregular dwarf probably represents an example of a dwarf group of extremely low luminosity. 
This extends the sequence of dwarf galaxy groups to the faint and ultra-faint luminosities. 

\begin{table*}
\centering
\caption{General parameters of LV\,J1157+5638 and LV\,J1157+5638\,sat}
\label{table:general}
\begin{tabular}{lcc}
\hline
                           &  LV\,J1157+5638          &  LV\,J1157+5638\,sat    \\
\hline
Position (J2000)$^a$       & $11^{\rm h}57^{\rm m}53.9^{\rm s} +\!56^\circ38'17''$  & $11^{\rm h}57^{\rm m}53.0^{\rm s} +\!56^\circ36'49''$ \\
$E(B-V)^b$, mag            & 0.017                    & 0.017                   \\
$V_T$, mag $^c$            & $16.61\pm0.04$           & $20.43\pm0.06$          \\
$I_T$, mag                 & $16.05\pm0.04$           & $19.75\pm0.06$          \\
$M_V$, mag                 & $-13.26\pm0.10$          & $-9.38\pm0.13$          \\
$M_I$, mag                 & $-13.80\pm0.10$          & $-10.04\pm0.13$         \\
Central surface brightness in $V$, mag arcsec$^{-2}$  & $21.15\pm0.04$ & $23.16\pm0.06$ \\
Central surface brightness in $I$, mag arcsec$^{-2}$  & $21.02\pm0.02$ & $22.71\pm0.06$ \\
Exponential scale length in $V$, arcsec & $3.23\pm0.02$ & $1.39\pm0.03$         \\
Exponential scale length in $I$, arcsec & $3.96\pm0.02$ & $1.45\pm0.04$         \\ 
Holmberg diameter in $V$, $a_{26.5}$, arcsec / kpc & 30.0 / 1.36   & 8.4 / 0.37 \\ 
Holmberg diameter in $I$, $a_{26.5}$, arcsec / kpc & 39.8 / 1.79  & 10.2 / 0.44 \\
Heliocentric radial velocity$^d$, \kms{}                    & $416.3\pm1.4$ & --    \\
Radial velocity relative to the Local Group$^e$, \kms{} & 514 & --              \\  
Distance modulus, mag                   & $29.82\pm0.09$   &  $29.76\pm0.11$    \\
Distance, Mpc                           & $9.22\pm0.38$    &  $8.95\pm0.42$     \\
Mean metallicity of RGB, $\FeH$, dex $^f$  & $-2.30\pm0.07$   & $-2.08\pm0.10$     \\ 
F(\ha), erg/cm$^2$ sec  $^g$            & $9.33\times10^{-14}$  & $<0.4\times10^{-14}$ \\
log(SFR)(\ha), $M_\odot$/yr             & $-2.10$          & $ < -3.50 $        \\                
m(FUV), mag$^h$                         & 18.49            & 22.71              \\
log(SFR)(FUV), $M_\odot$/yr             & $-2.63$          &  $-4.35$           \\
\hline
\end{tabular}
\begin{tabular}{p{0.8\textwidth}}
{\footnotesize{$^a$The measurements were made from the HST/ACS images.}} \\
{\footnotesize{$^b$From \citet{schlafly}}} \\
{\footnotesize{$^c$The total magnitudes and central surface brightness
are not corrected for Galactic extinction, whereas absolute magnitudes are corrected for
the Galactic extinction.}} \\
{\footnotesize{$^d$from SDSS DR12}} \\
{\footnotesize{$^e$from the Catalog \& Atlas of the LV galaxies database: http://www.sao.ru/lv/lvgdb/}} \\
{\footnotesize{$^f$The small [Fe/H] uncertainties are mostly reflect
the formal errors of the estimate defined by expression given in the Section 4.}} \\
{\footnotesize{$^g$ Using data from \citet{kar2015} and our distances}}\\
{\footnotesize{$^h$These GALEX magnitudes were obtained from the Mikulski 
Archive for Space Telescopes (MAST) (GALEX Public Release GR6/GR7). We estimate 
the respective SFR with the recipe given in the LV galaxies database for the similar data.}} \\
\hline
\end{tabular}
\end{table*}

\citet{weel2015} carried out hydrodynamic zoom-in simulations of isolated dark matter halos.
The authors demonstrate, that every halo is filled with subhalos, many of which form stars. 
The simulated dwarf galaxies with $M_*\simeq10^6 M_\odot$ host 1--2 satellites with $M_*=2{\rm -}200\times10^3 M_\odot$. 
There is the implication that dwarf galaxies throughout the universe should host tiny satellite 
galaxies of their own. 
\citet{dooley2017} also predict 1--6 (2--12) satellites with $M_* > 10^5 M_{\odot} 
(M_* > 10^4 M_{\odot})$ within the virial volume of LMC-sized galaxies, using
Caterpillar simulations. The authors emphasize an importance of finding and observing
of the faint satellites of dwarf galaxies for the determination of the
galaxy mass function and an importance of searches for faint dwarf groups,
which could test $\Lambda$CDM theory.

\section*{Acknowledgements}
This research is supported by award GO-13442 from the Space Telescope Science Institute
for the analysis of observations with Hubble Space Telescope.
This study is supported by the Russian Science Foundation (grant 14--12--00965).
We acknowledge the usage of the HyperLeda database (http://leda.univ-lyon1.fr).
GALEX data presented in this paper were obtained from the Mikulski Archive 
for Space Telescopes (MAST). STScI is operated by the Association of Universities 
for Research in Astronomy, Inc., under NASA contract NAS5-26555. Support for 
MAST for non-HST data is provided by the NASA Office of Space Science via 
grant NNX09AF08G and by other grants and contracts.
\bibliographystyle{mnras}
\bibliography{lvjtext1}   

\bsp
\label{lastpage}

\end{document}